# Fundamental Measurements in Economics and in the Theory of Consciousness

S. I. Melnyk[1,*] and I. G. Tuluzov[2]

[1]*Kharkov National University of Radio Electronics, Lenin Ave.4, 61161, Kharkov Ukraine.*

[2]*Kharkov Regional Centre for Investment, of.405, vul. Tobolska, 42a, Kharkov, 61072, Ukraine*

A new constructivist approach to modeling in economics and theory of consciousness is proposed. The state of elementary object is defined as a set of its measurable consumer properties. A proprietor's refusal or consent for the offered transaction is considered as a result of elementary economic measurement. Elementary (indivisible) technology, in which the object's consumer values are variable, in this case can be formalized as a generalized economic measurement. The algebra of such measurements has been constructed. It has been shown that in the general case the quantum-mechanical formalism of the theory of selective measurements is required for description of such conditions. The economic analogs of the elementary slit experiments in physics have been created. The proposed approach can be also used for consciousness modeling.

**Contents**

**Introduction**

Currently a large number of approaches to the description and modeling of economic systems exist. All of them are normally based on certain phenomenological assumptions on their properties and use classical (in the physical sense) methods of description of the system state. In this connection, the proposed models can be used only to the extent that the a priori assumptions on which they are based are correct. It contradicts to the general principle of model construction in theoretical and mathematical physics and does not allow using its full strength for the solution of problems of the economic theory.

We have earlier analyzed the possibility of developing a classical dynamics of the economic systems [1]. This analysis was not based on any physical analogs, and only used the methodology of the construction of dynamics established in exact sciences. It has been shown, in particular, that the law of surplus value in its simplest interpretation acts as an analog of the Newton's second law; besides such parameters of state of an economic system as mass and impulse have been determined, the conditions of occurrence of oscillatory motion have been analyzed, etc. This paper was awarded the Majorana Prize (EJTP) of 2010, as the best paper in the sphere of application of theoretical physics to the allied sciences.

The classical model is based on the assumptions on objective existence of exact values of such parameters of the system as its cost, profitability and their derivatives. These assumptions (the same as in classical physics) are valid only to the extent that the procedure of their measurements can be considered non-disturbing. Recently, an increasing number of works in the sphere of econophysics began to appear, in which the authors emphasize the necessity of refusing this approximation. The exact consecutive construction of the economic theory requires, the same as in other exact sciences, starting from the analysis of the measurement procedure. The present paper is dedicated to this specific subject.

Let us note that in the approach that we are going to develop in the present paper there are no principal differences between the real products and the virtual essences, which represent a value for humans in the framework of the theory of consciousness. Besides, a person's choice between the alternatives of behavior in the framework of such description turns out to be the analog of the notion of transaction. In this context, a "bargain with consciousness", for instance, can be described by the same mathematical symbols as a stock exchange transaction. Though we are further going to represent only the economic illustrations of the developed approach, it can be fully applied to the models of human consciousness as well, which is reflected in the title of the paper.

We have earlier analyzed the simplest procedure of measurement of state of a trader at an exchange, and it has been shown that for the description of its change as a result of various exchange events the quantum-mechanical formalism is required [2]. It is connected, in particular, with the phenomenon of a proprietor's consciousness, as an active element of the economic system. In a simplified form, we can say that the state of any economic system is characterized by the possibility of its participation in various economic events. At the same time this possibility depends both on the physical characteristics of the property itself (possibilities), and on the subjective properties of the proprietor's consciousness (his wishes and expectations). In the proposed approach, we are going to analyze the aggregate of the first and the second type of properties as a whole, as they have a joint effect on the result of the economic measurement. In this case, the quantum properties manifest themselves as an uncertainty in the trader's consciousness, which decreases in the process of decision-making (accepting or refusing a transaction) and changes the state of the whole analyzed system.

In the present paper we will formalize to the fullest possible extent the notions connected with the elementary measurement in the economic modeling and will develop the algebra of economic measurements (similarly to the algebra of selective measurements in the quantum mechanics).

1. **Determination of state of economic system. Elementary economic objects and elementary economic measurements**

There is a large number parameters of state of economic system discussed in the classical economic theory and in econometrics. However, most of them are secondary. They can be obtained on the basis of a priori economic and production parameters. Their objective existence is assumed, regardless of whether the system is being observed or not. At the same time, it is generally accepted that the fundamental science (as economics is claimed) must be based only on the measurable parameters of the system.

Therefore, before determining the notion of the state of economic system, let us introduce the primary notions of the elementary economic object and the elementary economic measurement. This will further help us to obtain the models of more complex systems exactly and sequentially, as it is performed, for instance, in fundamental physics.

The notion of "elementary" in science is referred not to some specific properties of objects, but to the logically consistent approach to their description. If the properties of any of the economic systems are determined in the final analysis by its consumer properties, then it is natural to call an **elementary economic object** such a system, which loses its individual consumer properties in case of division. This does not exclude the possibility of formation of new consumer properties of its parts, different from the initial properties. At this point it is appropriate to draw analogy with the elementary particles in physics, when each of the particles with sufficient energy can create a set of other particles. From this naturally follows the definition of **indistinguishable economic objects** as those having identical consumer properties. From now on, in order to avoid confusion, we are going to use the term "objects" for the economic systems having the same role as the physical particles in the acts of measurements of their state. Now for the completeness of the introduced definitions (in the mathematical sense), we must determine the notion of the consumer properties. It can be easily done by attaching them to the already formulated notions of the elementary economic objects. It is obvious that any consumer property is a certain qualitative and quantitative characteristic of the value of economic particle. And this means that it can be exchanged for any other economic objects having their own consumer properties. At the first glance, such definition seems non-constructive and resembles a "trick", which results in a "closed circle" of definitions, which lead to nowhere. But let's not be hasty in generalization. The same situation can be observed in any thesaurus. Certain words are interpreted using other words, and those others– using yet another word. And though the whole thesaurus is written in one language, and such predetermination results in a closed circle, we can eventually understand the essence of each of the words by studying their interconnections. The same situation can be observed in physics, though it is often "hidden under the carpet" of philosophy. The point is that the information about certain particles is transferred using other particles, information about those others – using yet another particle, etc. However, modern physics does not give the answer of how an exact and clear understanding of the situation is formed inside an observer's brain.

Therefore, the main postulate on which our model of economic measurements will be based is not the obvious procedure of redistribution of notions described above, but the following statement:

*The set of consumer properties defined as a possibility of exchange of certain elementary economic objects for other objects is <u>sufficient</u> for the description of the dynamics of economic systems.*

This means that even physically dissimilar objects having identical value in all possible exchanges are indistinguishable in terms of their consumer properties from the economic point of view. The above discussion allows us considering various product exchange procedures (transactions) as **the economic measurement**, as the consumer properties of any products are revealed as the result of this kind of transactions. Let us also note that not only the completed transactions, but also the transactions refused by one of the participants can be considered as the economic measurement, as both the consent and the refusal of the received offer contain information on the value of the elementary economic object. Therefore, the result of the economic measurement should be considered not only as a formal exchange procedure, but primarily as the information on the state of the elementary economic object provided by the result of the transaction.

Without discussing repeatedly the notion of the elementary character, let us define the *elementary*

*economic measurements* as a type of transactions, which cannot be half completed. In these transactions the subject of economy cannot accept part of the terms and conditions or trade. He either accepts the terms and conditions of the elementary transaction in full, or refuses.

The elementary economic event determined in this way is equivalent to the elementary measurements in physics. As Niels Bohr wrote, any physical measurement is based on the comparison with the etalon. And as a result of this comparison we can obtain only one of the two answers: positive or negative, while all other measurements can be represented as various combinations of elementary measurements. For this purpose the *algebra of economic measurements* must be developed, which will be performed in the present paper.

2. **Macro and micro objects in economics. Measurement using a classical measurer. Quantum properties of economic systems. Entangled states of elementary economic objects.**

Before proceeding to the development of the algebra of economic measurements, let us discuss their possible variants. In physics confusion sometimes occurs due to the fact that the notion of information obtaining does not have an accurate definition. For instance, in case of interaction of two particles the state of each of them changes in accordance with the change of state of the other. Is it possible to claim that in this case one particle measured the state of the other? Or, for instance, in the new interpretation of the "Schrodinger' cat paradox", proposed in [3], one of the observers sits in the same box as the cat. Has the measurement taken place if the inside observer already sees the new state of the cat and the outside observer still doesn't see it? The discussion of this and other methical issues is beyond the scope of this paper. Therefore, for the purpose of avoiding such confusions, we will divide the economic systems into macro- and microscopic. The interaction of elementary economic particle with the first type of systems will be referred to as the measurement and interaction with the second type - as the entangling of states, as it is established in modern quantum informatics. The economic essence of entangling will be discussed below.

We will define the **macroscopic economic system** as a complex system, in which the changing of state as a result of economic measurements (offer of transactions) can be neglected. Let us first note that in the real economy most systems are mesoscopic. It means that on the one side, they are sufficiently complex to be considered as elementary objects, and on the other side they are not sufficiently large to make it possible to neglect quantum effects in the process of observation of their state. We will discuss both idealized macroscopic economic systems (large companies, whose economic state remains practically unchanged as a result of transaction with one of the customers), and elementary economic objects (subject of economic activity, for whom the subject of transaction is all the property at his disposal).

Let us start with a more detailed analysis of measurements of the elementary economic particle using the macroscopic economic measure. The simplest example of such measurement is an offer to a proprietor to buy all his indivisible property at a certain price. As the notion of money is so far absent in our model, we will treat the purchase as an exchange for a certain quantity of units of product with other consumer properties. For instance, the employer offers the employee to purchase his working day for a certain quantity of grain. The further destiny of the employee depends on his decision to accept or to refuse the offer. He either becomes a participant of a certain technology (employs for the job) or becomes an independent economic particle. But are the consumer properties of his work force (considered as an elementary economic object) retained after such refusal?

Before answering this question, let us try to find an analog of such kind of elementary measurement in physics. In our opinion, the most successful illustration of this kind is the "screen experiment", widely used by Richard Feynman for the substantiation of his quantum-mechanical concept [4]. In our case we can consider that the equivalent of the transaction offered by the employer (payment for one working day in our example) is a certain semi-screen (Fig.1a) limiting from below the movement of a free particle. The vertical coordinate in the classical sense means the cost of one working day of a certain employee. If the employee accepts the transaction, the corresponding economic "particle" will be absorbed by the screen. In this case, its further destiny will depend on the properties of the screen and will not a subject of our current discussion. If the employee considers his cost higher, he will refuse the transaction and will continue his "flight". How will his state change in this case?

By obtaining the answer to this question we will practically describe the elementary act of economic dynamics in terms of the theory of measurements. For this purpose, we should keep in mind that the new state of an elementary economic particle is determined specifically by the possibility of participation of this employee in other economic events (transactions). It is practically obvious that, for instance, the employee who refused to exchange his working days for 5 measures of grain, will refuse to exchange it for 4 measures as well (the edge of the next "lower semi-screen" is lower than the previous one). But this obviousness decreases as the distance between the screens (time, during which the state of employee can change) is increasing. The answer to this question depends on many factors in both classical and quantum models, which cannot be completely taken into account. That is why, in the same manner as in physics, we can expect to predict the behavior not of an arbitrary multitude of employees, but only of the economic objects prepared in the same specific way. The economic essence of such "preparation" procedure will be discussed further.

In this paper we have used several times the term "quantum" for the description of properties of economic system. Now we can give a more precise definition of this term. It is often assumed in the process of discussing quantum properties that they are required for the description of dynamics of systems, which change their state as a result of measurement. However, it is also valid for a number of classical systems. Therefore we will refer as quantum to those economic systems, which change their state not only as a result of consent for a transaction (which is obvious), but also in case of refusal, thus accentuating the principal role of the information component of the description of state. Such properties can appear in economic systems only because the consciousness of the proprietor of certain products is an integral part of the system and to a significant degree determines its properties.

different products not according to their physical properties (quantity of proteins, fats, carbohydrates contained), but only according to the results of their economic measurements (which of course depend on their physical properties too).

In the classical model the measurements do not change the state of the observed object and their results do not depend on the order of measurement performance. Assuming that the measurements are of quantum nature, the result of the first of the measurements changes the а) The employer offers the price $S_k$ and «passes» for further measurements (offers) those applicants, who refused the offer. The lower semi-screen is the employer's obligation to provide work for the price $S_k$ б) The employer offers the price $S_i$ and «passes» for further measurements (offers) those applicants, who agreed for the offer. The upper semi-screen is the employee's consent to for the price $S_i$ в) The slit in the screen corresponds to the situation of the conflict of interests. The employer offers work for the price $S_k$, and the employee agrees to work for the price $S_i$. In case of refusing both offers the "particles" equivalent to the employees pass through the slit.

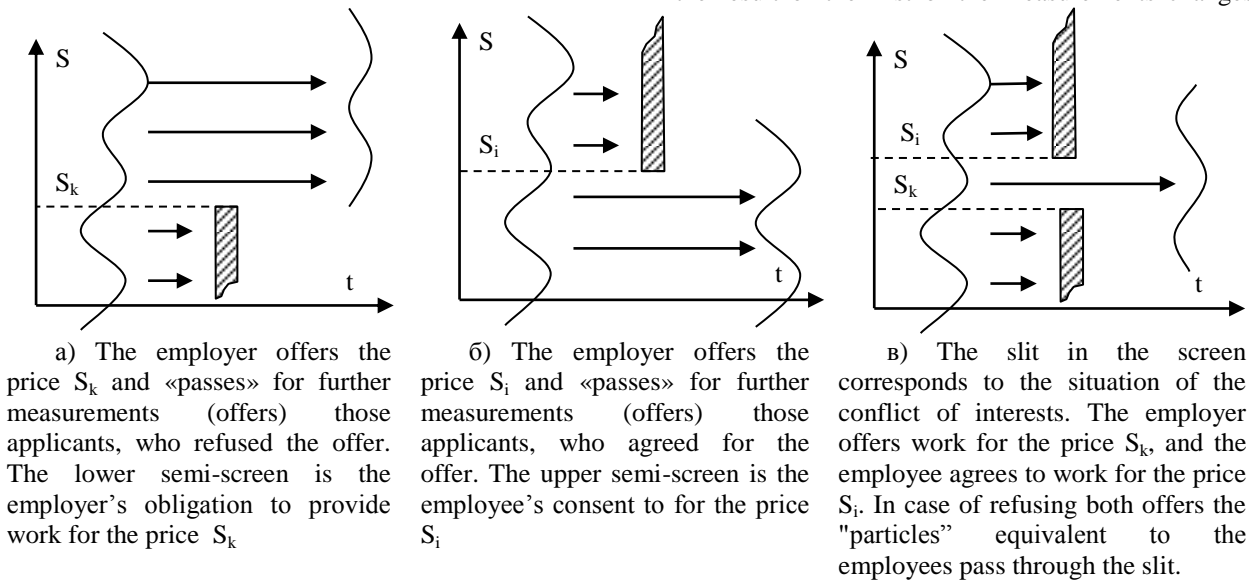

Fig.1. Slit analogs for the elementary economic measurements. The employer corresponds to the macroscopic classical object (semi-screen). The applicant for the vacancy corresponds to the elementary economic "particle", which either passes though the screen, or fails to pass.

Thus, by answering the question: "Where are the quantum properties of economic systems, which are being so much talked about?" we can state that they are contained in the consciousness of a proprietor who makes decisions in elementary economic events, thus determining the dynamics of the system. Let us note that some authors in physics are also inclined to attribute the quantum properties of the matter to the processes occurring primarily in the consciousness of observer.

Let us notice that the measurement procedure defined this way depends both on the product (grain in our example) of exchange, and on the conditions of the transaction, which can vary greatly. In the economic measurements they can be constructed arbitrarily be means of the corresponding rules of exchange. That is why a great number (even infinitely many) various economic measurements can exist. And if a development of a specific theory for each of such measurements was required, such theories would be no better than the phenomenological models used today. At the same time, we realize that the results of different economic measurements can be interconnected. For instance, the employee who refused an offer for 4 measures of grain is likely to refuse an offer for a similar quantity of other food containing the same amount of proteins, fats, carbohydrates and calories. In other words, the result of measurement "A" to a greater or lesser degree can determine the result of another measurement "B". The principal feature of our approach is that we compare state of the observed economic object (employee) and effects the result of the second measurement. Operators corresponding to such measurements do not commute. In physics the quantum nature of measurements is normally considered an integral fundamental property of the nature of matter. But in the economic systems we have linked them to the special features of perception and the phenomenon of choice. Therefore, in this case it is easier for us to find clear and usual illustrations of quantum effects.

Let us suppose, for instance, that in the initial (prepared) state the employee estimated his daily work for 6 measures of grain. If he is offered a price of 4 measures in the first measurement, he will probably refuse. But he can agree as well. Everything depends on the degree of his confidence of receiving a better offer in the future. In any case, if he will be offered 5 measures of grain after that, he will accept this offer with a higher degree of probability compared to the situation if there was no first measurement. It is because the first offer decreases his confidence of the price of his labor.

In quantum mechanics a similar phenomenon causes the quantum Zeno effect, the essence of which can be briefly described as the possibility of increasing the probability of a positive answer to the "question" by a sequence of "leading questions", changing the state of the quantum system. However, even the employee refuses both offers (passes both semi-screens) his state will depend on the sequence of the previously mentioned offers. Let us note that in both measurements the transactions are not performed, and the employee's state changes only as a result of the received information about the possible offers. Therefore, the quantum effects (connected to the non-commuting operators of measurements) primarily relate to the procedure of processing of the received information and to the solutions made as a result of this processing. For comparison, let us note that in the classical model all employees who had estimated their labor for 6 measures of grain will refuse both the first and the second offer with a guarantee.

From the physical point of view a human brain is a macroscopic classical object (though other hypotheses exist [5]). Therefore in principle any "Laplace's demon" with unlimited computing and time resources could model the procedure of human decision making, and by calculating all of its "hidden parameters", could predict the result of any economic measurement. Therefore, we state only the fact that in the process of describing the selection procedure as an indivisible elementary event the use of quantum-mechanical formalism is inevitably required. A rather rigorous mathematical argument of such necessity is given in the series of articles on the theory of quantum games [6,7], and an illustration (on the example of a "sea battle" game) is given in our paper [8].

So far, we have been discussing only the transactions between an elementary economic object and a macroscopic economic object (acting as a screen). However, most transactions in the real economy are performed between specific proprietors, none of which can be considered a macroscopic economic object. Such transactions (or refusals of such transactions) result in changing of state of both participants. The state of the first participant (offering exchange) is changed because his price turned out to be unacceptable, and the state of the second participant is changed because he was offered an unacceptable price (inadequate to his expectations).

In case if the offer would have been made by a macroscopic economic object (lower semi-screen with unknown position of the upper edge), then the object who had refused the transaction would have received exact information on the fact, that the cost of his property is not lower than the offered price. But if such transaction is offered by an individual vendor, it only means that he personally agrees to perform the offered exchange. Therefore, the result of such transaction (or refusal of it) becomes defined for one of the participants only after the economic state of the second participant will be measured by the macroscopic measurer. We can still consider the transaction a measurement in this case, but for the purpose of avoiding confusions, we will use the term "entangled states" as it is established in modern quantum informatics.

### 3. Algebra of economic measurements

For the construction of the mathematical apparatus of the theory of economic measurements we will use the methodology developed by J. Schwinger in the process of analysis of selective and non-selective measurements in physics. In his fundamental work [9] he has shown how to practically completely construct the whole quantum theory using only the general natural representations of measurements. This work is also attractive as the dynamics of the studied system is based only on the analysis of the properties of its changes, and is not based on the results of experiments like in other alternative interpretations. Therefore, our development will be also based from the very start only on the real values observed in economics (accept of refusal of the offered transaction by the subjects of economic relations), unlike the phenomenological econophysical models. We will develop the economic theory from the positions of positivism, considering the non-observable parameters only as patterns for the description of the real observed values. In this chapter we deliberately retain the terminology and designations proposed by Schwinger, but where required we will provide the economic interpretation of the symbols used in accordance with the aforesaid definitions. Though the rigorous mathematical construction used in [9] did not require the author to address to the properties of specific physical systems, we will propose physical analogs of the discussed economic illustration where it is considered appropriate.

The symbol of elementary measurement $M(a_i)$ (according to J. Schwinger) corresponds to the terms and conditions of the transaction, in which the employer (macroscopic economic system) offers a payment in the amount of $(a_i + \delta/2)$, and the applicant for the vacancy agrees for the payment $(a_i - \delta/2)$. It corresponds to the slit in the screen with the width δ. We introduce the **operation of addition** of symbols of economic measurements in such a way that the measurement

$$M(a_1) + M(a_2) = M(a_1 \cup a_2) \qquad (1)$$

«passes» only those proprietors, who assume that the consumer values of their property correspond to a wider interval $(a_1 \cup a_2)$. According to the results of this measurement it is impossible to determine to which of the two intervals they correspond.

But as each of the measurements $a_1$ and $a_2$ is set by a pair of obligations (those of the employee and the employer), their sum must be formed by a package of certain obligations. Now we will find out how these new obligations are connected with the initial obligations in case of addition of measurements. As each of the two obligations of the employer forms a certain lower semi-screen in our imagination, and the subject of addition are the slits, not the screen walls, then the superposition of the upper open semi-axes will result in the situation when only the lower of the two semi-screens will remain. The same rule applies to the employee's obligations, but in this case the higher of the two prices corresponding to his obligations will remain. The situation of two slits located close to each other is illustrated in Fig.2.

Even from this simple illustration follows the conclusion important for our further analysis – the number of employees "passing" through the sum of two slits cannot be calculated as a classical sum of employees "passing" through each of the slits separately. The point is that in the classical case the employee's opinion on the "adequate" payment of his labor does not depend on the specific nature of the offers he receives. But in the general case the situation is different.

Addressing to the physical analog, we can see that the particle passing over the edge of the semi-screen (a) may not pass over it if an edge of another semi-screen (a+δ) appears nearby. This effect is connected with the quantum properties of the particle and it is the stronger the narrower the slit between the semi-screens is. We can also state that the quantum properties become more significant compared to the classical properties when the width of the slit becomes equal to the de Broglies wavelength of the particle. On the basis of this analogy we can bind the **de Broglie wavelength of the economic particle** to the distance between the edges of the slit, which results in the violation of the classical law of addition of probabilities, or with the distance, at which the result of one measurement starts to effect the other.

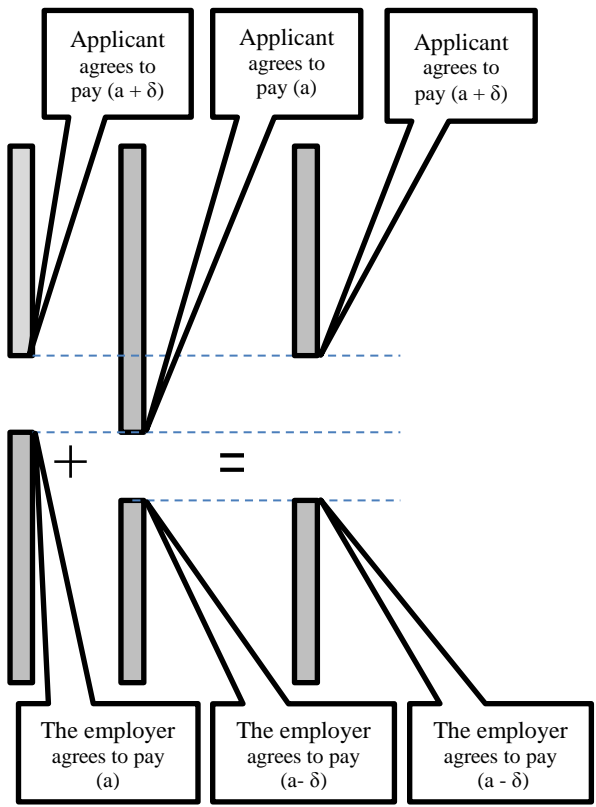

Fig.2. Addition of symbols of measurements results in the formation of a new package of obligations.

Let us note that unlike physics, the spectrum of values of wavelengths of elementary economic objects is continuous. It complicates the interpretation of various ensemble experiments. However, we can speak of specially prepared particles, as it is established in physics. Then we can consider the slit experiment (with variable slit width) as a method of measurement of the effective wavelength.

Let us introduce the **operation of multiplication** of the symbols of economic measurements as their sequential execution. It can be easily checked that for the above-introduced symbols of elementary economic measurements the operation of addition is commutative and associative, while the operation of multiplication is associative.

Let us also note that $M(a_i)M(a_i) = M(a_i)$. It means that a repeated offer of transaction with the same conditions will give the same result. Actually, this condition defines the completion of the measurement procedure as a performed choice and a corresponding change of state associated with this choice. Besides, for $a_j = \delta j$ we can write down for slit measurements:

$$M(a_i)M(a_j) = \delta(i,j)M(a_i), \qquad (2)$$

as at $i \neq j$ the result of one of the two semi-screen transactions of the first measurement contradicts to the transaction of the second measurement.

### 3.1. Compatible properties. Determination of economic state

All measurements defined above as $M(a_i)$, are homogeneous. They differ from each other only in the quantity of a certain economic value $(a)$, which is offered for exchange in the transaction. Along with this, we can also analyze other transactions in which the economic values $(b)$ are the subject of exchange. We will call the two economic measurements **compatible**, if the result of one of them does not depend on the result of the other. In this case, their order is negligible and thus we can write down the following:

$$M(a_i)M(b_j) = M(b_j)M(a_i) = M(a_ib_j) \qquad (3)$$

We will determine the **complete set** of economically compatible values as the maximum set of values, in which each pair is compatible. The economic state of the system is primarily determined by its consumer properties. And these properties appear (can be measured) as a result of various elementary economic events – transactions. Therefore the compatible consumer properties are those properties, which are not interconnected in any way at all. For instance, in case of using diamonds for strengthening the surface of cutting tools their transparency does not influence their price in any way. At the same time in case of using them for the production of optical elements their hardness is practically inessential. But such a quality (consumer value) of diamonds as the crystal size effects the price per carat both in the first and in the second case. Let us consider (very conventionally) that the diamond crystal size determines their consumer value in case of using them in jewelry. Then we can consider that the measurements of value of diamond carat in case of their "instrumental" and "optical" use are compatible, and the measurement in case of their "jewelry" use is not compatible either with the first or with the second measurement. Therefore, by finding out the price per diamond carat in instrumental industry (or receiving an offer of transaction) a customer can change his opinion on their "jewelry" value, but will not receive any information on their "optical" value. Let us emphasize that the term value refers, as previously, not so much to the physical properties of the diamonds, as to their consumer value, measured as a possibility of exchanging them for a certain product in a certain transaction. The received information cannot change their physical properties, but can change the state of consciousness of their owner and his estimate of their cost.

In physical models the equivalent of the compatible properties of quantum particles are the orthogonal vectors in a certain domain of state. In our next publication we will analyze the issue of the geometry of economic states in detail, and so far we will limit ourselves to the statement that both compatible and incompatible economic measurements exist. After determining the complete set of compatible economic parameters it is natural to introduce the notion of **complete economic measurement**. In our interpretation it is a set of transactions, the results of which determine the economic state of the system to the maximally complete degree. It means that no other transaction exists, the results of which do not depend on this state.

At this point we consider appropriate to quote a number of considerations of common nature, clarifying in a similar manner the essence of compatible and incompatible measurements both in economics and in physics. It follows from their determination that the result of any measurement, which is not included in the complete set, depends on the system state, which is determined by the results of the complete set of measurements. However, we state that an even stronger statement is valid: **the result of measurement not included in the complete set is completely determined by the results of the complete set of measurements.** The term completeness refers to the absence of any regularities not following from the results of the complete set of measurements.

Let us suppose that this statement is wrong. Then we can construct a complex measurement consisting of multiple iteration of the same testing, calculation of the statistical parameter characterizing the new regularity (not connected in any way to other results of the complete set of measurements) and forgetfulness (circulation) of the remaining information received in the process of testing. Such a composite measurement will be compatible with the rest of the measurements of the complete set. By adding it to the complete set we can ensure the validity of the stronger statement for all the remaining measurements.

### 3.2. Economic measurements changing the system states

Both in physics and in economics, it proves insufficient to use only the selective measurements of slit or semi-screen type for the description of the system dynamics. In the process of such economic measurements the only subject of changing is either the proprietary right for a certain product or the proprietor's opinion on its value. At the same time, the actual consumer properties change only as a result of utilization of these products. For commonality, we will refer to all such processes as the **technologies,** or measurements changing the state of the observed system (following the Schrödinger's terminology). Our aim is to show that these two types of measurements are sufficient for the description of the dynamic events in economic systems by the analogy with physics.

Let us note that in the "state-changing measurement" no selection of incoming elementary particles occurs. They all have different input and output states. Thus, the division of measurements into selective and state-changing allows representing all the observed dynamics as a structure consisting of such measurements. A question arises in this connection – to what extent this approximation is valid. In physics the approach based on the theory of continuous measurements is being increasingly used recently. In these measurements the processes of information reception, selection and changing of states occur continuously in time and parallel to each other. In our opinion, economic systems correspond to discrete description to a greater degree compared to physical systems. It is due to the fact that the exchange procedure is always documented by the transaction, while a continuously concluded transaction is nonsense. Nevertheless, in case if such measurements are performed sufficiently often, we can also replace their aggregate set by a single continuous measurement. This approximation has allowed us obtaining the generalization of the Black-Scholes formula for this specific case [10].

Measurements changing the system state are an integral part of any economic process. The exchange procedures (selective measurements) lose their sense if the resources received as a result of such exchanges will not be used for changing of state of their proprietor. In this context even an eaten hamburger is a technology changing the state of an elementary economic object (subject who ate this hamburger and who was adopting certain economic decisions).

Formally writing down the selective measurements $M(a')$ as the measurement $M(a', a')$, which changes the state for an identical, we can restrict ourselves to using only the second type of measurements for the description of the dynamics of economic systems. However, it can result in certain difficulties of logical nature. The point is that we associate with the selective measurement of the first type $M(a')$ only one transaction, in which the subject of economic measurement either accepts or refuses it. At the same time in the measurements of the type $M(b', c')$, changing the state, the objects in state (c') are selected at the input, while at the output they turn out to be in state (b'). In order to check it we can, for instance, prepare a set of objects in state (c') and make sure they all appear at the output. We can also pass the objects coming from the device through the slit measurement M(b') and make sure they all pass through the slit in the screen. Therefore, the equality $M(a') = M(a', a')$ means that the slit measurement $M(a')$ is a particular case of the technology, in which the state of object is not changed.

In the process of discussing economic measurements we will associate the measurements "at the input" and "at the output" of the measuring device with the transactions of two different types. The first type ("at the input") means that the proprietor receives advance payment, i.e.

before his state is changed. The second type ("at the output") assumes that the employee receives payment (for the results of his labor or other types of his property) upon completion of works. It is obvious that the first transaction characterizes his state before his participation in the technological process, and the second transaction - after it. In the general case, the results of these two measurements do not match. Correspondingly, the selective measurement $M(b',c')$ selects the employees, whose state is characterized by the economic slit (c') in case of advance payment for their work, or by slit (b') in case of payment upon completion of work.

### 3.3. Generalized form of notation of elementary economic measurements

Thus, any indivisible offer of transaction is an elementary economic measurement. At the same time each such offer is an obligation to perform certain terms and conditions of the transaction under conditions of the partner's consent to perform his part of obligations. However, most transactions are performed as a result of bargaining - mutual offers. Besides, the subject of agreement can be both advance payment and payment upon completion of works. In this connection we will introduce a generalized formula of measurement notation, in which all 4 offers are present. And all the rest of the transaction forms will be considered as its particular cases.

For the purpose of convenience we will use the symbolism taken from the quantum-mechanical formalism, denominating such measurement as $|A_2;B_2\rangle\langle A_1;B_1|$ or, in a simplified form - $|b_j\rangle\langle a_i|$. Such package of offers includes a pair of offers (obligations) from one participant (A – Alice) and a pair from the other (B - Bob), corresponding to the cost of payment for work performed (in our case, payment of the employee Bob paid by the employer Alice). Let us note that such form of notation is symmetrical both relative to the participants of the transaction and to its division for "input" and "output". In turn, Bob's work paid in advance can be considered as a payment for services performed by Alice paid upon their completion. In this form of notation the notions of "upper" and "lower" are also invariant for the semi-screens corresponding to each of the offers. For instance, obligation $B_1$ undertaken by Bob represents an upper semi-screen for him (he undertakes to perform the work if he is paid not less than $B_1$). If the transaction was not performed, it means Alice has rejected this offer. Thus, it acts as a semi-screen for her.

By making a pair of offers (for input and output), each of the participants of the transaction actually informs his or her partner on his agreement to perform them. Therefore, these offers are also the result of measurement. The complete result of the generalized economic measurement is thus determined by the whole aggregate of the offers made and by the consent or refusal to perform them by the partners.

As we have noted earlier, in the general case such measurement can be of local (only for two partners) nature. We will refer to it as the entangling of their states. In order to make sure that the information received as a result of measurement is generally accessible, one or both partners must be macroscopic. For this purpose, the measurement of their state as a result of performance or rejection of the transaction must be negligible. In the aforesaid slit experiments such "macroscopic" partner in the transaction acts as a screen, the state of which does not change regardless of whether the elementary economic object passed through the slit or didn't. Moreover, in this case both semi-screens forming the slit are macroscopic objects. In the economic interpretation it means that both offers (A and B) come from the same macroscopic partner – the employer. At the same time the second participant of the transaction, whose state is being measured, refuses one of the offers and accepts the other one while "passing through the slit". In the present paper we are further going to discuss only this type of measurements.

Let us note that in the framework of the generalized form of notation the differences between the technologies and the selective measurements (transactions) become inessential. Both in the first and in the second case the consumer properties of the economic objects are changed. And the question of whether it is connected with the material processes of processing of resources, or with the change of psychological estimate of their values is not so important.

### 3.4. Functions of transformation of economic measurements

Following the logics set forth in paper [9], we can consider the sequence of the technology $M(b',c')$ and the selective measurement of the technology $M(a')$ as a certain technology $M(a',c') \approx M(a')M(b',c') \equiv M(a',a')M(b',c')$, which selects the incoming particles in condition $c'$ and transforms them into the condition $a'$. At the same time, on the stage of selective measurement (selection) of particles with the property $b'$ through the "economic price slit" $a'$ only part of the particles pass through the slit. Therefore, we can conditionally write down

$$M(a')M(b',c') = \langle a'|b'\rangle M(b',c'), \quad (4)$$

where $\langle a'|b'\rangle$ is a certain factor why so far undetermined meaning. In the example discussed above it represents, for instance, those employees, who consider the cost of their working day equal to $b'$ measures of grain, but who can agree for a payment in the amount of $a'$ measures of meat as well. At the same time in the first case they refuse the offer $(b' - \delta_b/2)$, but agree for $(b' + \delta_b/2)$. And in the second case they refuse the offer $(a' - \delta_a/2)$, but accept $(a' + \delta_a/2)$.

Let us compare these employees with those, who were initially considering the cost of their working day (at the output) equal to $d'$ measures of vine, for instance. The aggregate measurement used for their selection can be written down as

$$M(a')M(d',c') = \langle a'|d'\rangle M(a',c'). \quad (5)$$

At the same time, it is possible that we can have the same number of employees in the same state $c'$ at the input of both measurements and the same number of employees, who had agreed that the cost of their working day is $a'$ at the output. But will the properties of these employees (relative to the subsequent measurements) be the same? After all, at the intermediate stage of selection the first agreed that the cost of their working days equals $b'$ measures of grain, and the latter agreed that cost of their working days equals $d'$ measures of vine. And if the answer is negative, then how will this information (about their intermediate choice) be reflected in the notation of the factor? Without answers to these questions the aforesaid equality is no more than a different form of notation, where the factor $\langle a'|d'\rangle$ is just a symbol denoting the selection of objects being in the state $d'$, which have subsequently passed through the slit M(a').

Making the formal calculation, we will take into account that $\sum_{a'} M(a') = 1$. This equality only means that the economic object of measurement will surely agree for the transaction for one of the possible prices. Then we can write down that

$$M(c',d') = \sum_{a',b'}\langle a'|c'\rangle\langle d'|b'\rangle M(a',b') \quad (6)$$

The economic meaning of the equality is that any technology $M(c',d')$ can be represented as the sum (by not as a mixture) of technologies $M(a',b')$ by going over various pairs $(a',b')$. By summarizing in (6) all the possible values of $(a',b')$ we are actually operating with packages of obligations corresponding to each of them. The sum of symbols of measurements, as above, only means the formation of a new package of offers from the summands according to specific rules. The term "sum of technologies" cannot be interpreted as a mixture of products of these technologies or as their simultaneous application to the same raw material (such explanation can prove to be completely impossible). The equality $M_1 = M_2 + M_3$ only denotes the rule, according to which we can calculate how the consumer properties of the object are changed as a result of measurement $M_1$, knowing how they are changed as a result of measurements $M_2$ and $M_3$. At the same time, let us note once again that the term "consumer properties" refers only to the measurable parameters in the fundamental economic measurements (possibility of consent or refusal from a certain transaction). Further, the same as in Schwinger's work, the **fundamental property of the composition of the transformation function** can be obtained:

$$\sum_{b'}\langle a'|b'\rangle\langle b'|c'\rangle = \langle a'|c'\rangle \quad (7)$$

Summarizing of various values of $b'$ formally means that in the interval between the selection of employees for $c'$ and $a'$ we are offering them to make a deal for the sale of their labor at any of the possible prices $b'$, which they are not even obliged to announce. Actually, we are passing through only the employees who agree that their labor has a certain cost in the units of the complete set $b$ as well. It is obvious that all employees will agree with this, and this consent will not change state and will not add any information about them. The mathematical essence of the fundamental quality is that if $a'$, $b'$ or $c'$ are the complete of compatible variables, then the laws of transformation of $a'$ into $b'$ and $b'$ into $c'$ are sufficient for calculating the laws of transformation of $a'$ into $c'$.

The formal analogy of this property with the laws of transformation between various reference systems in the classical mechanics can be noted. Actually, the results of the complete set of measurements represent the projections of the economic state of the set of objects on the corresponding reference system, and the set of symbols $\langle a'|b'\rangle$ represents the coupling coefficients of the descriptions of the same economic state in different reference systems.

It is worth making a remark on the **relative completeness** of the set of consumer properties described as $a'$, $b'$ or $c'$. It is determined only for the subset of consumer properties appearing in the discussed economic events (employment of workers). In case of taking account of a wider spectrum of properties connected with other transactions, these sets can turn out to be incomplete. At this point, the analogy with the measurements of electron spin projection in the experiments using the Stern-Gerlach device is appropriate. In case of measuring this projection in the plane perpendicular to the direction of movement of the electron, the complete set of selective measurements can be formed by any two perpendicular directions in this plane. At the same time, this set is incomplete relative to the measurements of the spin projection in all the directions of a three-dimensional space.

4. **Proceeding from the measurements in economic models to the theory of consciousness.**

In this paper, we have been so far discussing only the economic models and the possibility of application of the theory of selective measurements to them. However, the specifics of economic relations has been formalized by us in a rather general sense, as a change of consumer properties of economic elementary particles both as a result of transaction (or refusal of it) and in the process of a certain technological changing.

It is easy to note that the human behavior in the general case can also be described using these terms. We can consider a subject's consciousness as an elementary particle of consciousness. Then the selective measurement of state of such particle is a result of selection of a specific type of behavior under the specific circumstances. The set of these circumstances forms the terms and conditions of the transaction (with the corresponding "payments" depending on the adopted decision). If the subject's choice is limited to only one of the two alternatives – consent for a certain action (1) or refusal of it (0), then such measurement can be considered elementary.

By analogy with economics, among the infinite set of possible actions both compatible (decision on one of them is not connected in any way with the decision on the other action), and incompatible actions exist. The latter are immeasurably more numerous, but the circumstances determining them (terms and conditions of the transaction) can be expressed as a linear combination of compatible measurements of the complete set.

Let us note that at this level of description, the same as in economics, the subject's physical (or proprietary) capabilities to make a certain decision is an integral part of its state. In the framework of our theory of fundamental measurements, the subject's reluctance or incapability of performing a certain action is not important. The only result of the measurement is its final decision, which actually governs its behavior. The result of application of the theory of generalized selective measurements is the prediction of the statistical regularities of the decisions made by the subject based on the previous measurements. The processes occurring in his or her organism or mind are beyond the scope of the present theory.

Therefore, the "quantum" effects in human consciousness and behavior should be considered only in this meaning. We would like to stress that the "quantum nature of consciousness" in our context relates only to the semantics of its representation in the form of a sequence of elementary selective measurements. At the same time, we are considering only the completed processes of choice (transaction) and action (technology), without analyzing their possible mechanisms. In other words, we limit the set of measured parameters of state only to the obvious macroscopic events. The problems of quantum properties of consciousness analyzed by other authors relate rather to the perception of physical mechanisms of choice. The logic of our analysis does not change depending on whether they are quantum macroscopic events or classical events, as we are considering as the elements of consciousness the images and choices already formed by certain mechanisms, i.e. the classical results of brain activity. The necessity of applying the quantum-mechanical formalism only arises in case of an attempt to inscribe these results into the model of a certain enclosed system.

At the same time, the successfully developing theory of macroscopic quantum games [7] is the instrument capable of vividly representing the quantum properties of consciousness in the framework of the theory of selective measurements. So far, we have been considering the elementary measurement as an interaction of a classical macroscopic screen (e.g. a semi-screen) with a quantum elementary particle. The function of such a screen can be performed by an employer, whose state practically does not change regardless of the consent or refusal of the transaction of one or even of a certain number of employees. In this case, we can only consider that this macroscopic object is measuring the state of the particle in classical sense. However, if an interaction of two proprietors occurs, and each of them can be considered as an elementary economic object, the interaction between them can be considered as a bargaining, in which their roles are equivalent. At the same time, the result of such measurement (whether the transaction has taken place or not) characterizes the state of the pair of proprietors instead of the states of each of them separately. In terms of the modern theory of quantum information, such states are considered entangled. And only after an additional economic measurement of one of the participants of the transaction we can specify the state of the other participant.

Thus, for instance, if two subjects performed an exchange of a car for a house, we can only state on the basis of this fact that these two objects were of equal consumer value for them. However, in order to find out the consumer value for external observers we should perform an additional measurement of one of them. For instance, we should offer him a certain price for his house. Depending on his consent or refusal of such a transaction, we can specify the information on the "entangled" state of the other subject as well. The theory of quantum macroscopic games primarily deals with this type of entangled states. The connection of this theory with the possibilities of economic quantum modeling has been discussed by us in greater details in [8]. Let us note that the properties of compatibility of measurements for various subjects can be different as well. This does not result in contradictions in case of considering such subjects as "different particles".

Concerning the actual actions, they act as the technologies that change the consumer properties of elementary particles. By performing a certain action, the subject thus changes the circumstances and receives a possibility of participating in other transactions (or changes the probability of receiving a positive answer in certain elementary measurements). As a result, it turns out that the whole "line of behavior" of the subject can be described as a sequence of measurements and technologies (choices and actions). The task of dynamics in this case is obtaining of the equation (stochastic in the general case), which allows calculating the dynamics of measurement of the subject's state at the specified initial state and external influence. An example of such equation has been obtained by us in the process of analysis of the sequence of choices of a set of traders at an exchange, considered as a result of continuous fuzzy measurement of their quantum state [8]. The obtained system of equations is the quantum-mechanical generalization of the Black-Scholes formula and can be used for decreasing of economic risk.

**Conclusion**

Hereby we have shown that the general theory of generalized selective measurements can be used as a basis for the construction of economic models. At the same time, the states of economic systems can be formally determined as a set of their consumer properties considered as a possibility of exchange (transaction) or participation in a technology. The specific nature of transaction consisting in special properties of consciousness of proprietors depending on the offers they receive does not allow considering such states as classical and requires using the quantum-mechanical formalism.

In our opinion, the further development of the theory of economic measurements is seen in the construction of the space of states and research of its properties with account of symmetries specific for different types of transactions. Then the use of variational principles, such as the principle of risk minimization, the principle of absence of arbitration etc., will allow obtaining the equation of dynamics in the space of economic states.

The methods of such development of the theory are well-developed in physics and can be used as a basis for further calculations. In paper [9] it corresponds to the chapters "Geometry of states" and "Dynamical principle". At the same time, the same as in paper [1], no involvement of physical analogs and principles is required. All the necessary data can be and should be taken from the properties of the studied economic systems and from the formalized rules of performing of the generalized economic measurements.

On the basis of the proposed approach we have developed the illustrative model of the economic system corresponding to the double-slit experiment in physics. Due to the large volume of material, the analysis of the economic essence of this experiment is represented by us in the second part of this paper, finalized as a separate article.

______________________________________________________

[*]Electronic address: smelnyk@yandex.ru